\documentclass[11pt,a4paper]{article}
\usepackage{jheppub}

\usepackage{dsfont,hyperref}

\newcommand{\ud}{\mathrm d}
\newcommand{\reef}[1]{(\ref{#1})}
\newcommand{\beqa}{\begin{eqnarray}}
\newcommand{\eeqa}{\end{eqnarray}}
\newcommand{\nn}{\nonumber \\}

\title{Star Integrals, Convolutions and Simplices}

\author[a]{Dhritiman Nandan,}
\author[a,b]{Miguel F. Paulos,}
\author[a,b]{Marcus Spradlin}
\author[a,b]{and Anastasia Volovich}

\affiliation[a]{Department of Physics,
Brown University,
Box 1843,
Providence, RI 02912-1843,
USA}
\affiliation[b]{Theory Division,
Physics Department,
CERN,
1211 Geneva 23,
Switzerland}

\abstract{We explore single and multi-loop conformal integrals, such as the ones appearing in dual conformal theories in flat space. Using Mellin amplitudes, a large class of higher loop integrals can be written as simple integro-differential operators on star integrals: one-loop $n$-gon integrals in $n$ dimensions. These are known to be given by volumes of hyperbolic simplices. We explicitly compute the five-dimensional pentagon integral in full generality using Schl\"afli's formula. Then, as a first step to understanding higher loops, we use spline technology to construct explicitly the $6d$ hexagon and $8d$ octagon integrals in two-dimensional kinematics. The fully massive hexagon and octagon integrals are then related to the double box and triple box integrals respectively. We comment on the classes of functions needed to express these integrals in general kinematics, involving elliptic functions and beyond.}

\begin{document}
\maketitle

\section{Introduction}

The scattering amplitudes of $\mathcal{N}=4$ supersymmetric Yang-Mills theory~\citep{Brink:1976bc,Gliozzi:1976qd} with $SU(N)$ gauge group in the planar ($N\to \infty$) limit are remarkable objects, possessing many non-obvious properties. Chief among these are superconformal and dual superconformal symmetries~\citep{Drummond:2006rz,Drummond:2008vq} which close onto a larger group of Yangian symmetry~\citep{Drummond:2009fd,Drummond:2010qh,Beisert:2010gn,Drummond:2010uq}. Such symmetries together with on-shell recursions~\citep{Britto:2004ap,Britto:2005fq}, unitarity-based methods~\citep{Bern:1994zx,Bern:1994cg,Bern:2007ct}, the Grassmannian formulation of amplitudes~\citep{ArkaniHamed:2009dn,ArkaniHamed:2009vw,ArkaniHamed:2012nw} and the Wilson loop/scattering amplitude duality~\citep{Alday:2007hr,Brandhuber:2007yx,Drummond:2007au,CaronHuot:2010ek,Mason:2010yk} have greatly expanded our understanding of the $\mathcal N=4$ theory. These and other developments are reviewed in~\citep{SCATReview}.

One of the important outcomes of these ideas has been the tremendous progress in our knowledge about the structure of multi-loop amplitudes. Although the integrand of the theory has been completely constructed~\citep{CaronHuot:2010zt,ArkaniHamed:2010kv,Boels:2010nw,ArkaniHamed:2010gh}, new mathematical techniques are necessary to efficiently describe the integrated objects. One advancement along these lines has arisen from the study of the six particle Maximally Helicity Violating (MHV) amplitude or its remainder function at two loops~\citep{Bern:2008ap,Drummond:2008aq,DelDuca:2009au,DelDuca:2010zg}. Dual conformal symmetry, together with the proposed duality between scattering amplitudes and Wilson loops, enabled the analytic computation of this particular process, albeit at first only in a very complicated form. However, the knowledge about the possible space of functions of the scattering amplitude, which in this case were particular types of iterated integrals, and the application of the mathematical tool of symbols particularly suited to such integrals (see in particular~\citep{Goncharov:1994,Goncharov:1998,Goncharov:2002}), allowed for the simplification of this remainder function to a very manageable form~\citep{Goncharov:2010jf}. Symbol technology has seen several other important applications including~\citep{Gaiotto:2011dt,CaronHuot:2011ky,Duhr:2011zq,Brandhuber:2012vm,Duhr:2012fh}. When the number of kinematic variables is sufficiently small (in particular, for several six-particle processes), it has even been possible with sufficient effort to obtain analytic formulas for certain amplitudes (or Regge limits of amplitudes)~\citep{Dixon:2011pw,Dixon:2011nj,Dixon:2012yy,Drummond:2012bg,Pennington:2012zj}, but our analytic knowledge of more complicated integrals appearing in multi-loop SYM scattering amplitudes is still quite limited.

In a seemingly unrelated development it has been realized that the correlation functions of Conformal Field Theories (CFT's) in AdS/CFT at strong coupling have properties analogous to flat space scattering amplitudes in an auxiliary space called Mellin space, first introduced by Mack~\citep{Mack:2009mi,Mack:2009gy} and further studied in the works~\citep{Penedones:2010ue,Paulos:2011ie,Fitzpatrick:2011ia,Nandan:2011wc,Fitzpatrick:2011hu,Fitzpatrick:2011dm,Fitzpatrick:2012cg}. An application of this formalism in the context of flat space conformal integrals has appeared in~\citep{Paulos:2012nu}. In particular, it was shown that a large class of conformal integrals---including those corresponding to position space correlation functions in $\phi^4$ theory, which correspond to various kinds of box integrals---have a very simple Mellin representation which can be constructed in terms of Feynman rules. Using these, it is straightforward to see that there are simple integro-differential relations between various kinds of multi-loop integrals and lower loop ones, all the way down to a set of basic building blocks: the one-loop $n$-gon integrals in $n$ dimensions, also known as the $n$-point star integral. These relations generalize various differential relations between integrals of different loop order which have long been very useful in the study of scattering amplitudes (see in particular~\citep{Drummond:2010cz,Dixon:2011ng,Ferro:2012wa} for some recent examples relevant to SYM theory).

These results suggest that it is of pressing importance to understand the star integrals in detail (a close relative of our star integrals, with massless external legs but massive propagators, has been studied and evaluated explicitly in several cases in~\citep{Davydychev:1990jt}). In this note we take some modest steps in this direction. Firstly, it has been realized that such integrals compute volumes of simplices in hyperbolic space~\citep{Davydychev:1997wa,Mason:2010pg,Schnetz:2010pd,Paulos:2012qa} (a different relation between amplitudes and volumes has been explored in~\citep{ArkaniHamed:2010gg,ArkaniHamed:2012nw}). We can therefore use Schl\"afli's formula, which determines the differential of the volume of an $(n-1)$-simplex in terms of the volumes of $(n-3)$-simplices (a motivic version of Schl\"afli's formula~\citep{Goncharov:1996} has been similarly applied to compute symbols of star integrals in~\citep{Spradlin:2011wp}). As one application, we integrate the formula explicitly to find the $d=5$ pentagon integral. The result is remarkably simple, being simply a sum of logarithms with unit coefficients. The $d=6$ hexagon and $d=8$ octagon are addressed next. In these cases finding the full answer appears much more difficult (some special cases of the $d=6$ hexagon have been explicitly evaluated in~\citep{DelDuca:2011ne,Dixon:2011ng,DelDuca:2011jm,DelDuca:2011wh}) and we will content ourselves with finding analytic results when the external kinematics are restricted to two dimensions. We apply the results of the recently developed spline technology for loop integrals~\citep{Paulos:2012qa}, which tells us that in such kinematics, these integral can be written out as sums of box integrals with determined coefficients.

The fully massive $d=6$ hexagon ($d=8$ octagon) integral plays a role in determining the fully massive double (triple) box integrals in four dimensions. The relation of the $d=6$ hexagon to the double box has been worked out in~\citep{Paulos:2012nu}. In this note we do the same for the triple box and the octagon. We find the former is given as a double integral of the latter. Crucially, the hexagon and octagon integrals being integrated over are ratios of polylogarithm functions divided by certain square roots. We argue this has implications for the class of functions in terms of which higher loop integrals can be expressed.

The paper is organized in the following way. In Section~\ref{mellinsec} we review general ideas about Mellin amplitudes and the consequences of the existence of Feynman rules for Mellin amplitudes. Namely, we discuss the connections of multi-loop Feynman amplitudes with products of Mellin amplitudes, and its implications for the position space results. We stress that from this it is clear that we need to have a better understanding of $n$-gons in $n$ dimensions or the ``star" integrals to understand the fully massive loop integrals of $\mathcal{N}=4$ SYM. In Section~\ref{starsec} we discuss these star diagrams in more detail reviewing some known results as well as presenting some new analytic results for pentagons in five dimensions. For more complicated diagrams like the $d=6$ hexagon and the $d=8$ octagon it is very difficult to get explicit results for general kinematics. So, in Section~\ref{2dgonsec} we extensively discuss the analytic results for $2n$-gons using a restrictive kinematic localized in two dimensions. We do this by using the technology of splines to simplify such computations and present explicit results for two examples, the $d=6$ hexagon and the $d=8$ octagon. In Section~\ref{ellipticsec} we determine the representation of the triple box integral as a double integral of the $d=8$ octagon. Both in this case and for the double box, the integrand has a square root in the denominator which we know explicitly. We study various kinematic limits which tell us whether or not one should expect to see elliptic, or even more complicated, functions rather than the generalized polylogarithms which are much more familiar in multi-loop computations. Our results agree with the analysis of~\citep{CaronHuot:2012ab}. Some details about our results for the $d=6$ hexagon and the $d=8$ octagons in $2d$ kinematics are collected in the Appendices, and a Mathematica notebook with the full expressions is available in the online version of this note.

\section{Mellin amplitudes refresher}
\label{mellinsec}
\subsection{The Mellin amplitude}

The multi-dimensional Mellin transform formalism was introduced in the work of Mack~\citep{Mack:2009mi,Mack:2009gy} and quickly applied to both $AdS$/CFT~\citep{Paulos:2011ie,Fitzpatrick:2011ia,Nandan:2011wc,Fitzpatrick:2011hu,Fitzpatrick:2011dm,Fitzpatrick:2012cg} and flat space calculations~\citep{Paulos:2012nu,Costa:2012cb}. The Mellin transform can be applied to any conformally invariant function of several points $x_i$, with given conformal weights $\Delta_i$. This could be a conformally invariant correlation function or a conformally invariant integral (in applications to SYM theory scattering amplitudes, these are usually called {\em dual} conformal as a reminder that the relevant conformal symmetry is that in momentum space, rather than position space). For instance, we can write
\beqa
	\langle \phi_{\Delta_1}(x_1)\cdots \phi_{\Delta_n}(x_n)\rangle=\int [\ud \delta_{ij}]\, M(\delta_{ij})\prod_{i<j}^n \Gamma(\delta_{ij})\, x_{ij}^{-\delta_{ij}} \label{mellin}
\eeqa
where $x_{ij}\equiv (x_i-x_j)^2$ and the $\delta_{ij}$ parameters satisfy the constraints
	\beqa
	\sum_{i\neq j} \delta_{ij}=0, \qquad \delta_{ii}=-\Delta_i. \label{constraints}
	\eeqa
The function $M(\delta_{ij})$ is usually called the Mellin transform of $\langle \phi_{\Delta_1}(x_1)\cdots \phi_{\Delta_n}(x_n)\rangle$. After solving the constraints, the integral becomes an ordinary multi-variable Mellin transform in terms of $n(n-3)/2$ independent variables. The integration is over a set of complex variables $c_i$, each running from $-i\infty$ to $+i \infty$ along an appropriate contour. The constraints~\reef{constraints} guarantee that the variables $x_{ij}$ in the integrand combine into cross-ratios, thereby imposing conformality. It is important to note that the constraints can formally be solved by introducing a set of Mellin momenta $k_i$, satisfying momentum conservation, $\sum_i k_i=0$, such that
	\beqa
	\delta_{ij}=k_i\cdot k_j, \qquad k_i^2=-\Delta_i.
	\eeqa
This parameterization provides some intuition for the $\delta_{ij}$ parameters. In fact, in practice it is convenient to work with Mandelstam type variables, $s_{i_1\ldots i_p}=-(k_{i_1}+\ldots+k_{i_p})^2$, {\em e.g.} $s_{12}=-(k_1+k_2)^2=\Delta_1+\Delta_2-2 \delta_{12}$.

\subsection{Feynman rules and convolutions}
\label{convolutionsec}

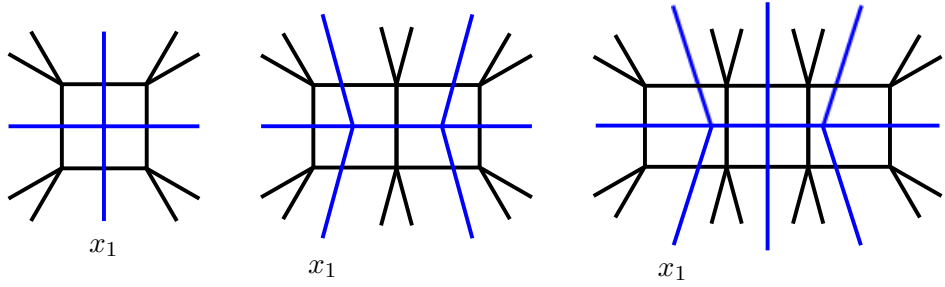
\begin{figure}
\begin{center}
\begin{picture}(350,85)
\put(0,0){\includegraphics[width=350pt]{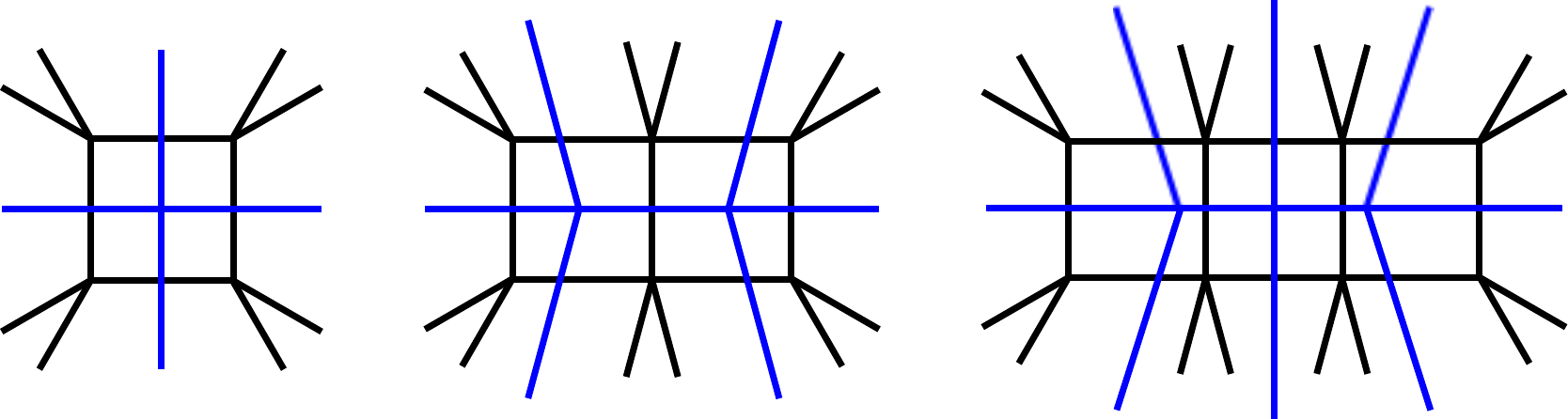}}
\put(36,1){\makebox(0,0){$x_1$}}
\put(118,-7){\makebox(0,0){$x_1$}}
\put(249,-8){\makebox(0,0){$x_1$}}
\end{picture}
\end{center}
\caption{The one-, two- and three-loop ladder diagrams (black) and their corresponding dual tree diagrams (blue). The external faces of the former, or equivalently the external vertices of the latter, are labeled $x_1,x_2,\ldots$ clockwise starting from $x_1$ as indicated.}
\label{boxfigure}
\end{figure}

In~\citep{Paulos:2012nu} a subset of us found that Mellin transforms of the kind of (dual conformally invariant) integrals that appear in SYM theory scattering amplitude computations have an extremely simple form. Consider for example a momentum space diagram whose position space dual is the same as a position space correlation function in $\phi^4$ theory (three examples are shown in Figure~\ref{boxfigure}, with the dual graphs shown in blue). The Mellin amplitude is obtained from the dual graph by the simple rules:
	\begin{itemize}
	\item To each external leg associate a Mellin momentum $k_i$ such that $k_i^2=-1$.
	\item Momentum flows through the diagram being conserved at each vertex.
	\item To each internal leg with momentum $k$ associate a propagator $1/(k^2+1)$.
	\end{itemize}
In other words, the Mellin amplitude looks just like a momentum space amplitude for massive $\phi^4$ theory, with $m^2=1$. This 1 is nothing but the canonical dimension of $\phi$, $\Delta=(d-2)/2=1$.

According to these rules we have, for example, the following very simple results for the Mellin amplitudes of the box, double box, and triple box integrals shown in Figure~\ref{boxfigure}:
	\beqa
	{\hbox{\lower 12pt\hbox{
	\begin{picture}(30,30)
	\put(0,0){\includegraphics[width=30pt]{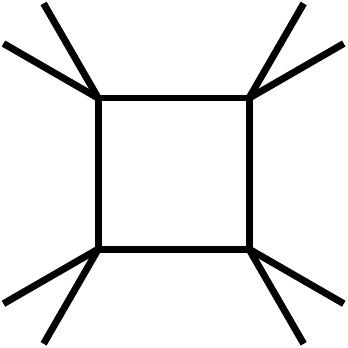}}
	\end{picture}}}} \qquad &\Longrightarrow& \qquad M=1, \\
	{\hbox{\lower 12pt\hbox{\begin{picture}(30,30)
	\put(0,0){\includegraphics[width=40pt]{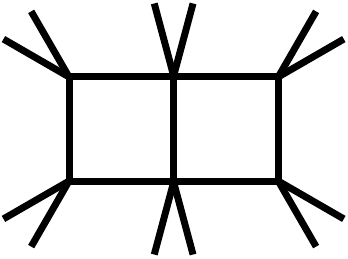}}
	\end{picture}}}} \qquad &\Longrightarrow& \qquad M=\frac{1}{1-s_{123}}, \\
	{\hbox{\lower 12pt\hbox{\begin{picture}(30,30)
	\put(0,0){\includegraphics[width=40pt]{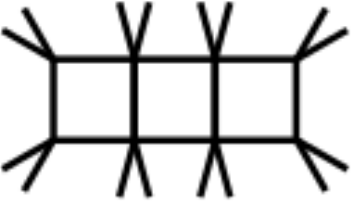}}
	\end{picture}}}} \qquad &\Longrightarrow& \qquad M=\frac{1}{1-s_{123}}\frac{1}{1-s_{567}}.
	\eeqa

\begin{figure}
\begin{center}
\begin{picture}(350,75)
\put(0,-20){\includegraphics[width=350pt]{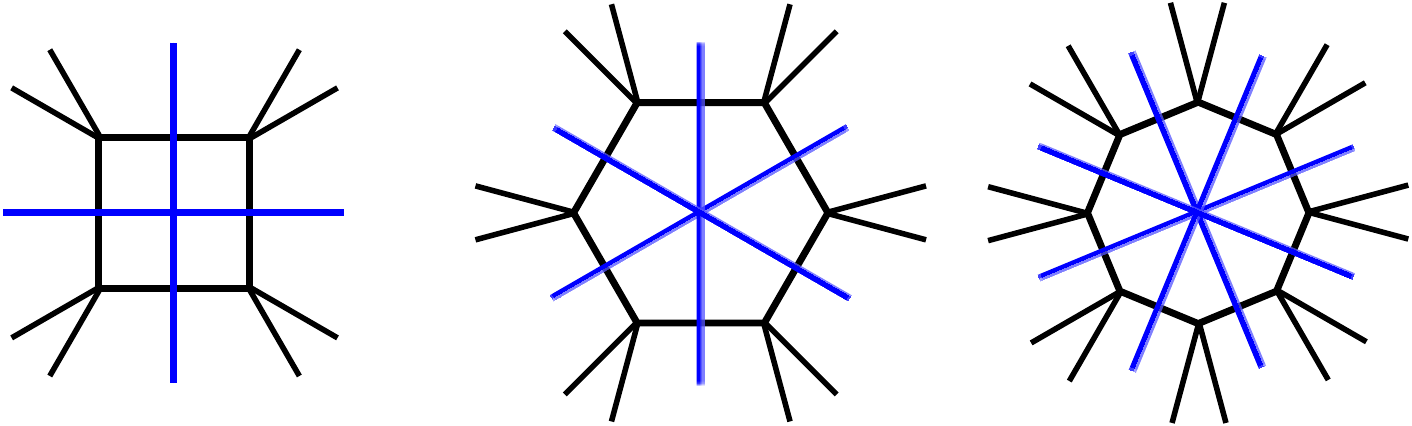}}
\end{picture}
\end{center}
\caption{The `star' graphs for $n=4,6,8$, in blue, correspond to the one-loop box, hexagon, and octagon integrals in $d=4,6,8$ respectively. These are the basic building blocks for many integrals relevant to multi-loop scattering amplitudes in SYM theory since each one is simply $M=1$ in Mellin space.}
\label{starfigure}
\end{figure}

The Feynman-like rules nicely express Mellin amplitudes as products of simple factors. We can use this to our advantage since a product in Mellin space maps back into position space as a convolution of the individual position space expressions. That is, suppose we have two functions $f(x), g(x)$ with Mellin transforms $M^f(s), M^g(s)$,
	\beqa
	M^f(s)=\int_0^{+\infty} \frac{\ud x}x\, x^{s}\, f(x), \qquad M^g(s)=\int_0^{+\infty} \frac{\ud x}x\, x^{s}\, g(x).
	\eeqa
Then the position space representation for the product $M^f(s) M^g(s)$ is
	\beqa
	h(x)&=&\oint \frac{\ud s}{2\pi i}\, M^f(s) M^g(s) x^{-s}=\oint \frac{\ud s}{2\pi i}\,
	\int_0^{+\infty} \frac{\ud y}y\, y^{s}\, f(y)M^g(s) x^{-s}\nonumber \\
	&=&\int_0^{+\infty} \frac{\ud y}y\, f(y) g(x/y). \label{piece1}
	\eeqa
Accordingly, we can split the computation of higher-loop integrals into two steps: first we compute the position space expression corresponding to the Mellin transform, which is just a product of propagators; and the second, more difficult step is to evaluate the position space expression of the product of $\Gamma$ functions appearing in~\reef{mellin}. But the latter is nothing but the same as computing a diagram whose Mellin amplitude is $M=1$, which corresponds to the $n$-legged star graph, examples of which are shown in Figure~\ref{starfigure}.

In SYM theory amplitude calculations we are also often interested in diagrams with various numerator factors. These can be translated into Mellin space as differential operators acting on the Mellin amplitude. Therefore we expect that a large class of integrals which appear in SYM theory scattering amplitude computations, to all loop order, can be expressed as integro-differential operators acting on just one class of elementary object: the $n$-point star integral in position space $\phi^n$ theory, or equivalently the one-loop $n$-gon Feynman integral in $n$ dimensions. This makes it clear that studying these objects is an important first step in understanding the analytic structure of a large class of multi-loop integrals.

\section{Star integrals}
\label{starsec}

It is convenient to use the embedding formalism~\citep{Dirac:1936fq,Weinberg:2010fx}. This amounts in practice to defining $d+2$-dimensional null vectors $P^M$ to describe $d$-dimensional coordinate vectors $x^\mu$, via
	\beqa
	P^M=(P^+,P^-,P^\mu)=(1,x^2,x^\mu).
	\eeqa
It is easy to check then that $P_{ij}\equiv -2 P_i\cdot P_j=(x_i-x_j)^2 = x_{ij}^2$.

The $n$-gon star integrals are defined by
	\beqa
	I^{(n)}=\int \frac{\ud^d x}{i\pi^{d/2}} \prod_{i=1}^n\frac{1}{(x_i-x)^2}=\int \frac{\ud^d Q}{i\pi^{d/2}} \prod_{i=1}^n\frac{1}{(-2 P_i\cdot Q)}.\label{starint}
	\eeqa
They are simply related to volumes $V^{(n-1)}$ of ideal hyperbolic $(n-1)$-simplices~\citep{Davydychev:1997wa,Mason:2010pg,Schnetz:2010pd,Paulos:2012qa} according to
	\beqa
	V^{(n-1)}=\frac{\sqrt{|\det P_{ij}|}}{2^{\frac{n}2} \Gamma\left(\frac n2\right)}\, I^{(n)}\label{simptoI}.
	\eeqa
Let us now consider the first few cases.

\subsection*{Triangle}

It is straightforward to do the integral directly in this case, and one finds
	\beqa
	I^{(3)}=\frac{\Gamma\left(\frac 12\right)^3}{\sqrt{P_{12}\,P_{13}\,P_{23}}}.
	\eeqa
Using formula~\reef{simptoI} above this gives $V^{(2)}=\pi$, which is indeed correct: the area of a hyperbolic ideal triangle is precisely equal to $\pi$.

\subsection*{Box}

The simplest non-trivial star integral is the first one in Figure~\ref{starfigure}, corresponding to the four-dimensional box function. The result for this well-known integral is given by
\begin{equation}
{\hbox{\lower 12pt\hbox{
\begin{picture}(30,30)
\put(0,0){\includegraphics[width=30pt]{1loopmassive.pdf}}
\end{picture}}}}
 = \frac{{\rm Li}_2(x_+/x_-) - {\rm Li}_2\left(\frac{1-x_+}{1-x_-}\right)
+ {\rm Li}_2\left(\frac{1-1/x_+}{1-1/x_-}\right) - (x_+ \leftrightarrow x_-)}{\sqrt{\det x_{ij}^2}}\label{boxint}
\end{equation}
in terms of
\begin{equation}
x_\pm = \frac{1}{2} \left( 1 + u_1 - u_2 \pm \sqrt{1 - 2 u_1 + u_1^2 - 2 u_2 - 2 u_1 u_2 + u_2^2} \right)
\end{equation}
and the two cross-ratios
\begin{equation}
u_1 = \frac{x_{13}^2 x_{24}^2}{x_{14}^2 x_{23}^2}, \qquad
u_2 = \frac{x_{12}^2 x_{34}^2}{x_{14}^2 x_{23}^2}.
\end{equation}
 The numerator in~\reef{boxint} is nothing but the Bloch-Wigner function (see {\em e.g.}~\citep{Schnetz:2010pd}), which indeed is known to compute the volume of an ideal hyperbolic tetrahedron.

\subsection*{Pentagon}

The next-simplest case, not shown in Figure~\ref{starfigure}, is the one-loop pentagon integral in five dimensions, which as far as we are aware has not been explicitly evaluated in the literature (the one-loop pentagon integral in four dimensions has been evaluated in~\citep{Bern:1993kr}). Surprisingly, we find that it takes a very simple form.

The pentagon integral corresponds to the volume of a hyperbolic 4-simplex. Such a volume depends on five cross-ratios, which in turn are built out of the five coordinates $x_i$. Let us take concretely
	\beqa
	u_1=\frac{P_{14}\,P_{23}}{P_{13}\, P_{24}}, \quad
	u_2=\frac{P_{25}\,P_{34}}{P_{24}\, P_{35}}, \quad
	u_3=\frac{P_{13}\,P_{45}}{P_{14}\, P_{35}}, \quad
	u_4=\frac{P_{15}\,P_{24}}{P_{14}\, P_{25}}, \quad
	u_5=\frac{P_{12}\,P_{35}}{P_{13}\, P_{25}}.
	\eeqa
To obtain an expression which only depends on cross-ratios we consider the rescaled integral
	\beqa
	\tilde I^{(5)}=\sqrt{P_{13}\,P_{14}\,P_{24}P_{25}P_{35}}\, I^{(5)}. \label{pentagonint}
	\eeqa
The computation of the volume is most straightforwardly done using Schl\"afli's formula. The formula relates the differential of a hyperbolic simplex in terms of its co-dimension 2 simplicial faces and associated angle differentials---since each co-dimension 2 face is defined by the intersection of two hyperplanes (which lie along co-dimension 1 faces), there is therefore an associated angle. This angle can be represented in terms of the vectors normal to said hyperplanes.

More concretely, if we have a simplex whose vertex representation is given by the $P_i$ vectors, its hyperplane representation is given in terms of vectors $W_i$ which are normal to these hyperplanes. In particular, $W_i\cdot P_j=\delta_{ij}$. In terms of these we can write Schl\"afli's formula as
	\beqa
	\ud V_{k}=\frac{-1}{2i (k-1)}\,\sum_{i<j}^n V_{(k-2)}^{(ij)}(-1)^{i+j} \,\ud \log\left(\frac{W_i\cdot W_j+\sqrt{(W_i\cdot W_j)^2-W_i^2 W_j^2}}{W_i\cdot W_j-\sqrt{(W_i\cdot W_j)^2-W_i^2 W_j^2}}\right)
	\eeqa
where $V_{(d-2)}^{(ij)}$ corresponds to the volume of the $d-2$ simplex spanned by all the $P_k$ vectors except for the pair $P_i, P_j$.

This formula is particularly simple in the case $k=4$. In this case the $V_{(k-2)}$ become volumes of ideal hyperbolic triangles. But this is simply $\pi$! The integration of Schl\"afli's formula depends on the kinematic region under consideration. We work in the Euclidean region where all $(x_i-x_j)^2$ are positive, and if we define
	\beqa
	\Delta^{(5)}&=&\frac 12 \frac{\det{P_{ij}}}{P_{13}\,P_{14}\,P_{24}P_{25}P_{35}} \nonumber \\
	&=& 1-[u_1 (1-u_3(1+u_4)+u_2 u_4^2)+\mbox{cyclic}]-u_1 u_2 u_3 u_4 u_5,
	\eeqa
then for $\Delta^{(5)}<0$ we have
	\beqa
	V_{(4)}=\frac{\pi }{6} \sum_{1\leq i<j\leq n} (-1)^{i+j}\log\left|\frac{W_i\cdot W_j-\sqrt{(W_i\cdot W_j)^2-W_i^2 W_j^2}}{W_i\cdot W_j+\sqrt{(W_i\cdot W_j)^2-W_i^2 W_j^2}}\right|,
	\eeqa
and using~\reef{simptoI} this gives
	\beqa
	\tilde I^{(5)}=\frac{\pi^{\frac 32}}{2 \sqrt{-\Delta^{(5)}}}\,\left(\sum_{1\leq i<j\leq 5} (-1)^{i+j}\log\left|\frac{W_i\cdot W_j-\sqrt{(W_i\cdot W_j)^2-W_i^2 W_j^2}}{W_i\cdot W_j+\sqrt{(W_i\cdot W_j)^2-W_i^2 W_j^2}}\right|\right).
	\eeqa
The pentagon has a cyclic permutation symmetry under the action of $g:u_i\to u_{i+1}$. We can then finally write the remarkably simple and manifestly symmetric form:
	\beqa
	\tilde I^{(5)}=\frac{\pi^{\frac 32}}{2 \sqrt{-\Delta^{(5)}}}(1+g+g^2+g^3+g^4)\left\{
	\log\left|\left(\frac{r-\sqrt{-\Delta^{(5)}}}{r+\sqrt{-\Delta^{(5)}}}\right)
	\left(\frac{s-\sqrt{-\Delta^{(5)}}}{s+\sqrt{-\Delta^{(5)}}}\right)\right|\right\}
	\label{pentagonresult}
	\eeqa
with
	\beqa
	r&=&\frac{(1-u_2)(1-u_5)-u_1(2-u_3-u_4-u_3 u_5-u_2 u_4+u_1 u_3 u_4)}2,
	\\
	s&=&\frac{(1-u_5)(1-u_2 u_5)-u_1\,(1+u_5-2\,u_3 u_5+u_4+u_2 u_4 u_5+u_1u_4)}{2\,\sqrt{u_1 u_5}}.
	\eeqa

\subsection*{Hexagon and beyond}

Using Schl\"afli's formula (see~\citep{Goncharov:1996,Spradlin:2011wp} for further details), one can easily express the differential (or, if one likes, the symbol) of the $n$-dimensional $n$-gon integral as a sum of certain $n-2$-dimensional $n-2$-gons. However, it is in general a difficult task to integrate this formula analytically. The structure of the differential equation makes it clear however that it can always be expressed in terms of generalized polylogarithm functions~\citep{Goncharov:1996}; in particular,
	\begin{itemize}
	\item $I^{(2n)}$ can be expressed in terms of functions of transcendentality degree $n$,
	\item and $I^{(2n+1)}$ can be expressed in terms of functions of degree
$n-1$.
	\end{itemize}
One way to understand the apparent inconsistency of the transcendentality counting in the two cases is that the odd-dimensional integrals always contain an overall factor of $\pi^{3/2}$, as we saw explicitly for the pentagon in~\reef{pentagonresult}. Taking this factor into account, the $m$-dimensional $m$-gon integral always has total degree $m/2$. We remind the reader that all generalized polylogarithms of degree less than 4 can be expressed in terms of the classical polylogarithms $\mbox{Li}_m$, so non-classical polylogarithms first appear in the $d=8$ octagon integral (for general kinematics).

We turn now to the $d=6$ hexagon integral, which has received attention in the literature~\citep{DelDuca:2011ne,Dixon:2011ng,DelDuca:2011jm,Spradlin:2011wp,DelDuca:2011wh} in part due to its interesting relationships (via differential equations) to other integrals relevant to SYM theory scattering amplitudes~\citep{Dixon:2011ng}. However it remains an interesting outstanding problem to fully evaluate the $d=6$ hexagon in general kinematics, where the integral depends on 9 independent cross-ratios (we present a choice of cross-ratios in Appendix~\ref{dbdetails}). To date the closest we have to this is the analytic formula for the special case of the ``three-mass easy'' hexagon~\citep{DelDuca:2011wh} (an expression for its symbol was given in~\citep{Spradlin:2011wp}). In this case three of the nine cross-ratios are set to zero. The formula presented in~\citep{DelDuca:2011wh} therefore computes the $d=6$ hexagon on a six-dimensional subspace of the full nine-dimensional cross-ratio space.

Motivated by the desire to simplify the evaluation of otherwise difficult integrals, and by the vast body of recent work on SYM theory amplitudes in two-dimensional kinematics (see for example~\citep{Heslop:2010kq,Alday:2010jz,Heslop:2011hv,Ferro:2012wa,Goddard:2012cx}), in this paper we therefore carry out explicit computations of the $d=6$ hexagon and the $d=8$ octagon in $2d$ kinematics. Here, due to Gram determinant constraints, the nine cross-ratios for the hexagon (and the twenty cross-ratios for the general octagon) are constrained to take values in a six-dimensional (ten-dimensional) subspace of the full parameter space. We present explicit parameterizations of the cross-ratios in terms of six (ten) free variables in Appendices~\ref{dbdetails} and~\ref{tbdetails}. Our result for the $d=6$ hexagon in $2d$ kinematics is in a sense complementary to that of~\citep{DelDuca:2011wh} since the two six-dimensional subspaces are disjoint inside the full nine-dimensional parameter space of the generic $d=6$ hexagon.

\section{$2n$-gon loop integrals in $2d$ kinematics}
\label{2dgonsec}

\subsection{Setup: splines}

In this section we evaluate $2n$-gon loop integrals in two-dimensional kinematics. To do this we shall use the methods developed recently in~\citep{Paulos:2012qa} based on spline technology, which the reader should consult for further details. With these, it can be shown that the one-loop star integral~\reef{starint} can be written in the form
	\beqa
	I^{(n)}=2\,\int_{\mathds M^D} e^{X^2}\, \mathcal T(X;\{P_i\})\label{i2n}
	\eeqa
where the spline is defined by
	\beqa
	\mathcal T(X;\{P_i\})=\int_0^{+\infty} \prod_{i=1}^{n} \ud t_i\, \delta^{(D)}(X-\sum_{i=1}^{n} t_i P_i).
	\eeqa
This expression follows by noticing that the spline is the Laplace transform of the integrand. Here we are interested in $2d$ kinematics, so we set $D=d+2=4$. We shall also only consider even-dimension integrals and therefore set $n\to 2n$. The computation of the spline depends on the various linear relationships between the $P_i$'s. Here we shall assume that the vectors are generic, i.e.~that every set of four vectors spans $\mathds M^4$.

Under these conditions the spline can be written as a sum of terms, each corresponding to a particular linearly independent set of vectors. Not all such sets need be considered though. It is sufficient to take the set $\mathcal B$ of so-called unbroken basis, which for generic kinematics amounts to the set of basis which include the vector $P_1$. To each such basis, $b$, there corresponds a piece in the spline, which is therefore made up of $N=(n-1)!/(n-4)!3!$ terms. Each term is labeled by its unbroken basis, $b$, and the coefficients can also be easily computed. In this manner we find
	\beqa
	\mathcal T(X;\{P_i\})=\sum_{b\in \mathcal B}\, \frac{\left(W_1^{(b)}\cdot X\right)^{2n-4}}{\prod_{i=1}^{2n-4}\,W_1^{(b)}\cdot \hat P_i^{(b)}}\, \frac{\chi_{(b)}(X)}{\sqrt{\det b^T b}}.
	\eeqa
Some explanations are in order. Firstly, $\hat P_i^{(b)}$ denotes the $i$th vector {\em not} in the basis $b$. Secondly the vectors $W_i^{(b)}$ are defined by
	\beqa
	W_i^{(b)}\cdot P_j=\delta_{ij},\quad \forall P_j\in b.
	\eeqa
We can think of $b$ itself as a matrix whose columns are the vectors $P_i\in b$. This allows us to compute the determinant. Finally, $\chi_{(b)}(X)$ is the characteristic function of the cone spanned by the vectors in $b$, which can be written as
	\beqa
	\chi_{(b)}(X)=\prod_{i=1}^4 \Theta(W^{(b)}_i \cdot X).
	\eeqa
To proceed we must evaluate the Gaussian-type integral in~\reef{i2n}. We could evaluate it directly, since the spline is homogeneous in $|X|= \sqrt{-X^2}$. This would give us a sum of integrals of $X$ polynomials over $AdS$ tetrahedra. However, instead of doing this we can use the presence of the exponential to integrate by parts the terms of the form $W\cdot X$. At the end of this procedure, there are no such factors left, but there are however several types of terms, depending on how many times we differentiate the characteristic functions $\chi_{(b)}(X)$. In particular, one set of terms does not involve derivatives of at all:
	\beqa	I^{(n)}&=&\frac{(n-4)!!}{2^{\frac{n}2-2}}\,\sum_{b\in \mathcal B}\, \frac{\left(W_1^{(b)}\right)^{n-4}}{\prod_{i=1}^{n-4}\,W_1^{(b)}\cdot \hat P_i^{(b)}}\int_{\mathds M^4} e^{X^2}\,
	\, \frac{\chi_{(b)}(X)}{\sqrt{\det b^T b}}+\ldots\ .
	\eeqa
This is interesting, as the integrals above are nothing but box integrals, with four external legs $P_i$ corresponding to the elements in the basis $b$. Accordingly, the kind of terms above are simply a sum of box integrals, namely dilogarithms. In contrast, the $\ldots$ represent terms which have an even number of derivatives of $\chi_{(b)}(X)$. We have explicitly checked that all such terms cancel between themselves for $n=3,4$. To understand why, notice that those terms involve for example integrals over lines in $AdS$, which leads to single logarithmic terms. In order to have an expression of uniform transcendentality, it must be that these terms actually add up to zero.

\subsection{Applications: hexagon, octagon, and beyond}

To see in detail how we can perform the computation of these coefficients, let us set $n=3$ and consider the particular basis made up of elements $P_1,P_2,P_3,P_4$. We then have
	\beqa
	W_1^{(1234),M}=\frac{\epsilon^M_{\ NPQ} P_2^N P_3^P P_4^Q}{\epsilon_{ABCD} P_1^A P_2^B P_3^C P_4^D}\nonumber \\
	\Rightarrow \frac{ \left(W_1^{(1234)}\right)^2}
	{\left(W_1^{(1234)}\cdot P_5\right)\left(W_1^{(1234)}\cdot P_6\right)}&=&\frac{\delta^{MNP}_{ABC} P_{2,M} P_{3,N} P_{4,P} P_2^A P_3^B P_4^C}{
	\delta^{MNPQ}_{ABCD}\,P_{5,M}P_{2,N} P_{3,P} P_{4,Q} P_6^{A} P_2^B P_3^C P_4^D}
	\eeqa
with $\delta^{A_1\ldots A_N}_{B_1\ldots B_N}$ the totally antisymmetric product of $N$ delta functions. It is important to notice that this expression, when multiplied by the inverse of $\sqrt{\det b^T b}$, will have total homogeneity $-1$ in each of the vectors $P_i$. Although we have focused on a particular term, this is a generic feature. It guarantees that, if we multiply $I_{2n}$ by $P_{14}P_{25}P_{36}$, each term in the sum is separately conformally invariant, and can hence be written in terms of the nine cross-ratios of a conformal six point function (though we must keep in mind the result is only valid for $2d$ kinematics, which imposes non-linear relations on these cross-ratios).
We give a choice for these in Appendix~\ref{dbdetails}, together with a $2d$ kinematics parameterization for them in terms of 6 independent variables $\chi_i^{\pm}, i=1,2,3$. In terms of the latter, we can write the contribution of the particular basis (1234) to $I_{6}$ as
	\beqa
	I_{6}&=&\frac{(2n-4)!!}{2^{n-2}}\,
	\frac{\chi ^-_1 \chi ^-_2 \chi ^+_1 }{ \left(\left(\chi
   ^-_1-\chi ^-_2\right) \chi ^+_1+\left(\chi ^-_1+1\right) \chi ^-_2 \chi ^+_3\right) \left(-\chi ^+_1+\chi ^+_3+\chi ^-_1 \left(\chi
   ^+_3+1\right)\right)}\times\nonumber \\
   &&\times\frac{\left(\chi ^-_1+1\right)\left(\chi ^+_1-\chi ^+_3\right) \left(\chi ^+_3+1\right)^2}{\left(\chi ^+_3 \left(\chi ^+_3-\chi ^+_1\right)+\chi ^-_1 \left(\left(\chi ^+\right)_3^2+\chi ^+_1-\chi ^+_2 \left(\chi
   ^+_3+1\right)\right)\right)}\times B+\ldots,
    \nonumber \\
    B&=&2\, \text{Li}_2\left(\frac{\chi ^+_1-\chi ^+_3}{\chi ^+_2-\chi ^+_3}\right)+2\, \text{Li}_2\left(\frac{\chi ^-_1-\chi ^+_3}{\chi ^+_3 \chi ^-_1+\chi
   ^-_1}\right)+\nonumber \\
   &&\log \left(\frac{\chi ^-_1 \left(\chi ^+_1-\chi ^+_3\right) \left(\chi ^+_3+1\right)}{\left(\chi ^-_1-\chi ^+_3\right) \left(\chi
   ^+_2-\chi ^+_3\right)}\right) \log \left(-\frac{\chi ^-_1 \left(\chi ^+_1-\chi ^+_2\right) \left(\chi ^+_3+1\right)}{\left(\chi ^-_1+1\right)
   \left(\chi ^+_2-\chi ^+_3\right) \chi ^+_3}\right)+\nonumber \\
   &&\log \left(\frac{\chi ^+_3-\chi ^-_1}{\chi ^-_1 \left(\chi ^+_3+1\right)}\right) \log
   \left(\frac{\chi ^+_3-\chi ^+_1}{\chi ^+_2-\chi ^+_3}\right)+\frac{\pi ^2}{3}
   \eeqa
Overall, there are a total of ten such terms. The total result is too cumbersome to reproduce here, but in the online version of this note we include a Mathematica notebook with the full result.

The computation of the $d=8$ octagon integral in $2d$ kinematics is entirely analogous to what we have just done. There are now a total of 35 terms in the spline, each corresponding to a box integral with a certain coefficient. The $d=8$ octagon depends on 20 cross-ratios which in $2d$ kinematics can be parameterized in terms of 10 independent parameters. The details of this kinematics have been included in Appendix~\ref{tbdetails}. The full expression for $I^{(8)}$ for the $d=8$ octagon have been included in the attached Mathematica file since it is very lengthy.

It is straightforward to consider generalizations of the results above and consider $2n$-dimensional integrals in $2m$ kinematics, for $n>m+1$. Under such circumstances one finds the $2n$-dimensional integral decomposes (for generic $2m$-dimensional kinematics) into a sum of $(2n-1)!/(2n-2m)!(2m-1)!$ $2m$-integrals with well defined coefficients. For instance, the general even-dimensional integral in $4d$ kinematics is given by
	\beqa	I^{(2n)}&=&\frac{(2n-6)!!}{2^{n-2}}\,\sum_{b\in \mathcal B}\, \frac{\left(W_1^{(b)}\right)^{2n-6}}{\prod_{i=1}^{n-6}\,W_1^{(b)}\cdot \hat P_i^{(b)}}\int_{\mathds M^6} e^{X^2}\,
	\, \frac{\chi_{(b)}(X)}{\sqrt{\det b^T b}}+\ldots.
	\eeqa
For the $d=8$ octagon the number of unbroken basis made up of six vectors is 21 and accordingly the $d=8$ octagon is a sum of 21 $d=6$ hexagon integrals.

\section{Elliptic functions and beyond}
\label{ellipticsec}

\subsection{The double box}

One of the motivations for this work was to make an attempt to begin exploring integrals which evaluate to functions outside the class of generalized polylogarithm functions. Elliptic functions of this type have been encountered before in explicit QCD computations~\citep{Laporta:2004rb}, and have been argued to appear in SYM theory as well starting with a double box integral contribution to the 2-loop 10-point N${}^3$MHV amplitude~\citep{CaronHuot:2012ab}.

Using the convolution tricks explained in Section~\ref{convolutionsec}, it was shown in~\citep{Paulos:2012nu} that the 3 to 3 exchange diagram in position space of $\phi^4$ theory, which is the same as the double box Feynman integral, can be expressed as a one-fold integral of the 6-point star (the $d=6$ hexagon integral):
\beqa
I_{3,3}(u_1,\ldots,u_8,u_9)= \int_{u_8}^{+\infty}\frac{\ud u_8'}{u_8'}\, \tilde I^{(6)}(u_1,\ldots,u_8',u_9). \label{doublebox}
\eeqa
with $\tilde I^{(6)}=x_{14}^2 x_{25}^2 x_{36}^2 I^{(6)}$ and the double box integral,
\beqa
I_{3,3}=\int \frac{\ud^4 x_a \ud^4 x_b}{(i\pi^2)}\frac{x_{14}^2 x_{25}^2 x_{36}^2}{x_{1a}^2 x_{2a}^2 x_{3a}^2 x_{4b}^2 x_{5b}^2 x_{6b}^2\,x_{ab}^2}.
\eeqa
Thanks to our results in Section~\ref{2dgonsec} we are now in possession of a simple formula giving the $d=6$ hexagon in $2d$ kinematics. One therefore may hope
that this should suffice for determining the double box in the same kinematical regime. However the formula above demands that the integration is done keeping all cross-ratios fixed except one, and it is easy to check that this is impossible in $2d$ kinematics, since the number of independent cross-ratios in this case is reduced because of Gram determinant identities. It is somewhat unfortunate that in order to recover a lower-dimensional kinematics result we have to take a detour through the full, generic result. Similar remarks hold for higher loop integrals: although we are only interested in $4d$ kinematics at the end of the day, our convolution formulae nevertheless require a detour through a higher-dimensional regime.

The symbol of the fully general $d=6$ hexagon is known~\citep{Goncharov:1996,Spradlin:2011wp}, but it is rather complicated, and integrating it in general remains an interesting open problem. Since obtaining the full result seems to be currently out of reach, what can we say about it? Well, firstly we know what form the final expression has to take. We know that the $d=6$ hexagon integral is related to the volume of a 5-simplex living in an $AdS_5$ submanifold of $AdS_7$. Denoting this volume by $V_5$, we have from formula~\reef{simptoI} (and neglecting numerical factors):
	\beqa
	I^{(6)}(x_i)\simeq \frac{V_5}{\sqrt{\det x_{ij}^2}}. \label{hexaratio}
	\eeqa
Schl\"afli's formula tells us that the differential volume of the 5-simplex is fixed entirely in terms of that of the 3-simplex, and from this we know that the result will take the form
	\beqa
	I^{(6)}(x_i)\simeq \frac{ \mbox{Li}_3(\ldots)+\ldots}{\sqrt{\det x_{ij}^2}},
\eeqa
where in the numerator of course $\mbox{Li}_3()$ is shorthand for various terms of the correct transcendentality, such as $\mbox{Li}_2()\log(), \log()\log()\log()$ and $\zeta(3)$, with complicated functions of the cross-ratios as arguments.

Our expression for the double box integral then becomes
	\beqa
 I_{3,3}(u_i)=\int_{u_8}^{+\infty}\frac{\ud u_8'}{u_8'}\, \frac{ \mbox{Li}_3(\ldots)+\ldots}{\sqrt{\Delta^{(6)}}}. \label{dbox}
	\eeqa
with $\Delta^{(6)}=\frac{\det x_{ij}^2}{(x_{14}^2 x_{25}^2 x_{36}^2)^2}$.
In general, $\Delta^{(6)}$ is a third-order polynomial in $u_8$, 
	\beqa
	\Delta^{(6)}= \left[4\, u_1 u_2 u_5 u_6 u_7  u_9 u_8^3 + \mbox{lower-order terms in } u_8 \right].
	\eeqa
Therefore, if any three cross-ratios are set to zero (and both $u_3$ and $u_4$ must be included in the three), then the determinant necessarily reduces to a second-order polynomial in $u_8$. This is important, since the order of the polynomial determines whether we should expect elliptic functions to appear in the final expression for the double box after integrating~\reef{dbox}. Indeed, if we get rid of the polylogarithms for a second, the integral
	\beqa
	\int \frac{\ud u_8}{u_8 \sqrt{(u_8-a)(u_8-b)(u_8-c)}}
	\label{sqrt}
	\eeqa
leads to elliptic functions for generic $a,b,c$. If any pair of roots degenerates, or if the polynomial becomes second order instead of cubic, we would obtain logarithms instead. Because of this, it seems almost certain that the final integrated expression for the double box will contain elliptic functions, in general kinematics.

Let us look at a particular limit of the general kinematics where we actually expect to start seeing the elliptic functions in the final result. For the $d=6$ hexagon the ``minimal massive" case where we go beyond polylogarithms would be the case of $4$ massive legs. Say we have $x^2_{61} = 0$ and $x^2_{34} = 0$. In this case we have $u_3 = u_4 = 0$, and as argued above this is the largest number of vanishing cross-ratios we can have while staying within the realm of elliptic functions. This configuration is exactly the case appropriate to the 10-point double box integral shown in Figure~$6$ of~\citep{CaronHuot:2012ab}. Now if we further set the other cross-ratios (apart from $u_8$) to some constant, generic values, then~\eqref{sqrt} certainly gives an elliptic function, so we would expect the same to be true for the double box in~\reef{dbox}. However any other case with a smaller number of massive legs only gives polylogarithms, never elliptic functions, because for such cases the polynomial inside the square root degenerates from cubic to at most quadratic order. It might be interesting (and certainly easier) to derive the hexagon integral with $u_8$ arbitrary, $u_3=u_4=0$ and all other cross-ratios set to some carefully chosen kinematic values. This would be sufficient to plug into formula (\ref{dbox}) and check if elliptic functions actually do occur there.

\subsection{The triple box}

Let us now consider the triple box integral. In the dual position space this looks like a tree-level diagram involving 8 particles and two internal propagators (shown in Figure~\ref{boxfigure}). Accordingly we expect it to be given by a two-fold integral of the 8-point star integral. This is what we shall proceed to show just now.

To begin with, we need a basis of cross-ratios which can describe a conformally invariant function of $8$ points. For fully generic kinematics (in general dimension) we expect $8\times 5/2=20$ independent cross-ratios. We list a choice of such cross-ratios in Appendix~\ref{crossratios}. Next, we consider the Mellin representation of the triple box. According to the rules we set out in Section~\ref{mellinsec} it is the product of two propagators. Once the constraints~\reef{constraints} are solved, we get an ordinary multi-dimensional transform in terms of the 20 independent cross-ratios. In this way we find
	\beqa
	I_{3,2,3}=\int_{-i\infty}^{+i\infty}\left(\prod_{i=1}^{20} \frac{\ud c_i}{2\pi i}\, u_i^{c_i}\right)\, \frac{1}{(2+2 c_{12})(2+2 c_9)} \times \prod_{i<j}^8 \Gamma(\delta_{ij}). \label{tripleboxone}
	\eeqa
with
	\beqa
	I_{3,2,3}=\frac 14 \int \frac{\ud x_a \ud x_b \ud x_c}{(i\pi)^3}\, \frac{x_{15}^2 x_{26}^2 x_{37}^2 x_{48}^2}{x_{1a}^2\,x_{2a}^2\, x_{3a}^2\, x_{4b}^2\,x_{5b}^2\, x_{6c}^2\,x_{7c}^2\,x_{8c}^2\, x_{ab}^2\,x_{bc}^2}
	\eeqa

For the reader's benefit we provide in Appendix~\ref{gammas} an explicit formula for the product of 28 $\Gamma$-functions written out in terms of 20 independent variables $c_i$.

We don't need to display all those details here since we already know that the position space expression corresponding to the product of gamma functions is nothing but the $d=8$ octagon integral. We have therefore only to compute the (much simpler) position space expressions corresponding to the propagator factors. For instance,
	\beqa
	\int_{-i \infty}^{+i \infty}\, \frac{\ud c_{12}}{2\pi i}\, \frac{ u_{12}^{c_{12}}}{2(1+c_{12})}= \frac{\Theta(1-u_{12})}{2\,u_{12}}
	\eeqa
with $\Theta(x)=1$ for $x>0$, and zero otherwise. In this manner we conclude that
	\beqa
	I_{3,2,3}(u_1,\ldots,u_{20})=\frac{1}{u_9\,u_{12}} \int_{u_9}^{+\infty}\ud u_9'\int_{u_{12}}^{+\infty}\,  \ud u_{12}'\, \tilde I^{(8)}(u_1,\ldots,u_9',\ldots,u_{12}',\ldots,u_{20}). \label{triplebox}
\eeqa
with $\tilde I^{(8)}=x_{15}^2x_{26}^2 x_{37}^2 x_{48}^2 I^{(8)}$.
Of course, this equation can be turned around to write a (very simple) differential equation expressing the octagon as a second derivative of the triple box.

Finally let us remark on the $d=8$ octagon integral. It is again given by the volume of a hyperbolic simplex, and accordingly we have something of the schmatic form
\beqa
	\tilde I^{(8)}(u_i)=\frac{ \mbox{Li}_4(\ldots)+\ldots}{\sqrt{\Delta^{(8)}}}.
\eeqa
with $\Delta^{(8)}$ now given by $\frac{\det{x_{ij}^2}}{(x_{15}^2x_{26}^2x_{37}^2x_{48}^2)^2}$.
Of course we emphasize that the numerator will be a linear combination of (generalized) polylogarithm functions of degree four, including not just $\mbox{Li}_4()$ but also for example $\mbox{Li}_{2,2}()$, $\mbox{Li}_2() \mbox{Li}_2()$, etc.
If we evaluate the determinant, we find
	\beqa
\Delta^{(8)}=\frac{u_1 u_2^3 u_3 u_4^2 u_5^2 u_6 u_7^2\,u_8^3 u_{10} u_{11}^3}{u_{13}\,u_{14}\,u_{16}^2\, u_{18}\,u_{19}}\,(u_9^3 u_{12}^3+\ldots)
	\eeqa
where the $\ldots$ stands for terms of lower degree in $u_9$ or $u_{12}$. For general kinematics the first integration with respect to $u_9$ will make elliptic functions appear, while it is reasonable to expect that (again, in general kinematics) the second integration will lead to an even more complicated class of functions, beginning as we see at three loops.

\acknowledgments

We are grateful to S.~Caron-Huot for stimulating discussions.
This work was supported by the US Department of Energy under contracts
DE-FG02-91ER40688 (MP, MS) and DE-FG02-11ER41742 (AV Early Career Award),
the Simons Fellowship in Theoretical Physics (AV),
and the
Sloan Research Foundation (AV).  MP also acknowledges funding from the
LPTHE, Paris, and is grateful to the CERN for hospitality during the
course of this work.

\appendix

\section{Details on the double box computation}
\label{dbdetails}

A list of 9 multiplicatively independent cross-ratios required to describe conformally invariant functions of six point is given by the following set:

\beqa
u_1=\frac{x^2_{14} x^2_{23}}{x^2_{13} x^2_{24}},\quad u_2=\frac{x^2_{15} x^2_{24}}{x^2_{14} x^2_{25}},\quad u_3=\frac{x^2_{16} x^2_{25}}{x^2_{15} x^2_{26}},\quad u_4=\frac{x^2_{25}x^2_{34}}{x^2_{24} x^2_{35}},\nonumber \\
u_5=\frac{x^2_{26} x^2_{35}}{x^2_{25} x^2_{36}},\quad u_6=\frac{x^2_{12} x^2_{36}}{x^2_{13} x^2_{26}},\quad u_7=\frac{x^2_{36} x^2_{45}}{x^2_{35}
   x^2_{46}},\quad u_8=\frac{x^2_{13} x^2_{46}}{x^2_{14} x^2_{36}},\quad u_9=\frac{x^2_{14} x^2_{56}}{x^2_{15} x^2_{46}}.
   \label{uhex}
\eeqa

In order to carry out the computation for the $d=6$ hexagon in $2d$ kinematics we can first restrict the general kinematics of~\eqref{uhex} to a four-dimensional sub-space parameterized by $12$ momentum twistors~\citep{Hodges:2009hk}. Subsequently, we can further restrict the $4d$ momentum twistors to a subspace of $2d$ kinematics which can be very simply parameterized using $6$ independent cross-ratios, as a generalization of the parameterization used in~\citep{Alday:2010jz,Ferro:2012wa,Goddard:2012cx}:
\begin{eqnarray}
   Z_{1}&=&\left(  \begin{array}{c}   0 \\   0 \\   i \sqrt{2} \chi_2^+ \\   \frac{i (1-\chi_2^+)}{\sqrt{2}} \\  \end{array}  \right),~~~
       Z_{2}=\left(  \begin{array}{c}  i \sqrt{2} \chi_3^- \\  \frac{i (1-\chi_3^-)}{\sqrt{2}} \\  0 \\  0 \\  \end{array}  \right),~~~
     Z_{3}=\left(  \begin{array}{c}   0 \\   0 \\   i \sqrt{2} \chi_3^+ \\   \frac{i (1-\chi_3^+)}{\sqrt{2}} \\  \end{array}  \right),~~~
  Z_{4}=\left(  \begin{array}{c}  i \sqrt{2} \chi_1^- \\  \frac{i (1-\chi_1^-)}{\sqrt{2}} \\  0 \\  0 \\  \end{array}  \right).\nn
      Z_{5}&=&\left(  \begin{array}{c}   0 \\   0 \\   i \sqrt{2} \chi_1^+ \\   \frac{i (1-\chi_1^+)}{\sqrt{2}} \\  \end{array}  \right),~~~
      Z_6=\left(    \begin{array}{c}  0 \\  \frac{i}{\sqrt{2}} \\   0 \\     0 \\   \end{array}  \right),~~~
      Z_7=\left( \begin{array}{c}  0 \\  0 \\  i \sqrt{2} \\    - i \sqrt{2} \\  \end{array}  \right),~~~
Z_8=\left(  \begin{array}{c}   -i \sqrt{2} \\  \frac{i}{\sqrt{2}} \\    0 \\  0 \\    \end{array}   \right),\nn
 Z_9&=&\left( \begin{array}{c}  0 \\  0 \\  i \sqrt{2} \\    - i \sqrt{2} \\  \end{array}  \right),~~~
Z_{10}=\left(  \begin{array}{c}   -i \sqrt{2} \\  \frac{i}{\sqrt{2}} \\    0 \\  0 \\    \end{array}   \right),~~~
   Z_{11}=\left(  \begin{array}{c}  0 \\ 0 \\   0 \\  \frac{i}{\sqrt{2}} \\   \end{array}  \right),~~~
      Z_{12}=\left(  \begin{array}{c}  i \sqrt{2} \chi_2^- \\  \frac{i (1-\chi_2^-)}{\sqrt{2}} \\  0 \\  0 \\  \end{array}  \right).
 \label{12ptkin}
\end{eqnarray}
In terms of these variables one may compute the $x_{ij}^2 = \det(Z_{2i-1} Z_{2i} Z_{2j-1} Z_{2j})$, so that the $9$ cross-ratios~\eqref{uhex} are given by
 \beqa
  u_1&=&\frac{\chi _1{}^- \left(\chi _3{}^-+1\right)
   \left(\chi _1{}^+-\chi _3{}^+\right)}{\left(\chi
   _1{}^-+1\right) \chi _3{}^- \left(\chi
   _1{}^+-\chi _2{}^+\right)}, \nn
 u_2&=&\frac{\left(\chi _1{}^-+1\right) \left(\chi
   _2{}^++1\right)}{\left(\chi _3{}^-+1\right) \left(\chi
   _3{}^++1\right)}, \nonumber \\
 u_3&=&\frac{\left(\chi _2{}^--\chi _3{}^-\right) \chi
   _2{}^+ \left(\chi _3{}^++1\right)}{\left(\chi
   _2{}^--\chi _1{}^-\right) \left(\chi
   _2{}^++1\right) \chi _3{}^+}, \nonumber \\
 u_4&=&\frac{\chi _3{}^++1}{\chi _1{}^+ \chi
   _1{}^-+\chi _1{}^-+\chi _1{}^++1}, \nonumber \\
 u_5&=&\frac{\left(\chi _2{}^--\chi _1{}^-\right) \left(\chi
   _1{}^++1\right) \chi _3{}^+}{\chi _2{}^- \chi
   _1{}^+ \left(\chi _3{}^++1\right)}, \nonumber \\
 u_6&=&\frac{\chi _2{}^- \left(\chi _1{}^--\chi
   _3{}^-\right) \chi _1{}^+ \left(\chi
   _2{}^+-\chi _3{}^+\right)}{\left(\chi
   _2{}^--\chi _1{}^-\right) \chi _3{}^-
   \left(\chi _1{}^+-\chi _2{}^+\right) \chi
   _3{}^+}, \nonumber \\
 u_7&=&\frac{\chi _2{}^- \chi _1{}^+}{\left(\chi
   _2{}^-+1\right) \left(\chi _1{}^++1\right)},\nonumber  \\
 u_8&=&\frac{\left(\chi _2{}^-+1\right) \chi _3{}^-
   \left(\chi _1{}^+-\chi _2{}^+\right)}{\chi
   _2{}^- \left(\chi _3{}^-+1\right) \chi _1{}^+}, \nonumber \\
 u_9&=&\frac{\chi _3{}^-+1}{\chi _2{}^+ \chi
   _2{}^-+\chi _2{}^-+\chi _2{}^++1} .
   \label{uhex2d}
\eeqa

\section{Details on the triple box computation}
\label{tbdetails}

\subsection{Cross-ratios for eight-point functions}
\label{crossratios}

A list of 20 multiplicatively independent cross-ratios required to describe conformally invariant functions of eight points is given by the following set:
	\beqa
u_1&=&\frac{x_{15}^2 x_{24}^2}{x_{14}^2 x_{25}^2},\quad u_2=\frac{x_{16}^2
   x_{25}^2}{x_{15}^2 x_{26}^2},\quad u_3=\frac{x_{17}^2 x_{26}^2}{x_{16}^2
   x_{27}^2}, \quad u_4=\frac{x_{26}^2 x_{35}^2}{x_{25}^2 x_{36}^2},\nonumber \\
   u_5&=& \frac{x_{27}^2
   x_{36}^2}{x_{26}^2 x_{37}^2},\quad u_6=\frac{x_{28}^2 x_{37}^2}{x_{27}^2
   x_{38}^2},\quad u_7=\frac{x_{37}^2 x_{46}^2}{x_{36}^2 x_{47}^2},\quad u_8=\frac{x_{38}^2
   x_{47}^2}{x_{37}^2 x_{48}^2},\nonumber \\ u_9&=&\frac{x_{13}^2 x_{48}^2}{x_{14}^2
   x_{38}^2},\quad u_{10}=\frac{x_{48}^2 x_{57}^2}{x_{47}^2
   x_{58}^2},\quad u_{11}=\frac{x_{14}^2 x_{58}^2}{x_{15}^2
   x_{48}^2},\quad u_{12}=\frac{x_{15}^2 x_{68}^2}{x_{16}^2
   x_{58}^2},\nonumber \\ u_{13}&=&\frac{x_{13}^2 x_{28}^2}{x_{12}^2
   x_{38}^2},\quad u_{14}=\frac{x_{13}^2 x_{24}^2}{x_{14}^2
   x_{23}^2},\quad u_{15}=\frac{x_{24}^2 x_{35}^2}{x_{25}^2
   x_{34}^2},\quad u_{16}=\frac{x_{35}^2 x_{46}^2}{x_{36}^2
   x_{45}^2},\nonumber \\ u_{17}&=&\frac{x_{46}^2 x_{57}^2}{x_{47}^2
   x_{56}^2},\quad u_{18}=\frac{x_{57}^2 x_{68}^2}{x_{58}^2
   x_{67}^2},\quad u_{19}=\frac{x_{17}^2 x_{68}^2}{x_{16}^2
   x_{78}^2},\quad u_{20}=\frac{x_{17}^2 x_{28}^2}{x_{18}^2 x_{27}^2}.
   \label{uocta}
   \eeqa
As in the previous section for the double box computation in $2d$ kinematics we can use the momentum twistor parameterization of the above cross-ratios in terms of 16 momentum twistors in a four-dimensional subspace, which are again expressed in a $2d$ subspace parameterized by $10$ cross-ratios. The momentum twistor representation is given by,
\begin{eqnarray}
 Z_1&=&\left( \begin{array}{c}  0 \\  0 \\  i \sqrt{2} \\    - i \sqrt{2} \\  \end{array}  \right),~~~
Z_{2}=\left(  \begin{array}{c}   -i \sqrt{2} \\  \frac{i}{\sqrt{2}} \\    0 \\  0 \\    \end{array}   \right),~~~
   Z_{3}=\left(  \begin{array}{c}  0 \\ 0 \\   0 \\  \frac{i}{\sqrt{2}} \\   \end{array}  \right),~~~
      Z_{4}=\left(  \begin{array}{c}  i \sqrt{2} \chi_1^- \\  \frac{i (1-\chi_1^-)}{\sqrt{2}} \\  0 \\  0 \\  \end{array}  \right),\nn
       Z_{5}&=&\left(  \begin{array}{c}   0 \\   0 \\   i \sqrt{2} \chi_1^+ \\   \frac{i (1-\chi_1^+)}{\sqrt{2}} \\  \end{array}  \right),~~~
      Z_6=\left(    \begin{array}{c}  0 \\  \frac{i}{\sqrt{2}} \\   0 \\     0 \\   \end{array}  \right),~~~
      Z_7=\left( \begin{array}{c}  0 \\  0 \\  i \sqrt{2} \\    - i \sqrt{2} \\  \end{array}  \right),~~~
Z_8=\left(  \begin{array}{c}   -i \sqrt{2} \\  \frac{i}{\sqrt{2}} \\    0 \\  0 \\    \end{array}   \right),\nn
      Z_{9}&=&\left(  \begin{array}{c}   0 \\   0 \\   i \sqrt{2} \chi_2^+ \\   \frac{i (1-\chi_2^+)}{\sqrt{2}} \\  \end{array}  \right),~~~
       Z_{10}=\left(  \begin{array}{c}  i \sqrt{2} \chi_2^- \\  \frac{i (1-\chi_2^-)}{\sqrt{2}} \\  0 \\  0 \\  \end{array}  \right),~~~
     Z_{11}=\left(  \begin{array}{c}   0 \\   0 \\   i \sqrt{2} \chi_3^+ \\   \frac{i (1-\chi_3^+)}{\sqrt{2}} \\  \end{array}  \right),~~~
  Z_{12}=\left(  \begin{array}{c}  i \sqrt{2} \chi_3^- \\  \frac{i (1-\chi_3^-)}{\sqrt{2}} \\  0 \\  0 \\  \end{array}  \right),\nn
      Z_{13}&=&\left(  \begin{array}{c}   0 \\   0 \\   i \sqrt{2} \chi_4^+ \\   \frac{i (1-\chi_4^+)}{\sqrt{2}} \\  \end{array}  \right),~~~
       Z_{14}=\left(  \begin{array}{c}  i \sqrt{2} \chi_4^- \\  \frac{i (1-\chi_4^-)}{\sqrt{2}} \\  0 \\  0 \\  \end{array}  \right),~~~
     Z_{15}=\left(  \begin{array}{c}   0 \\   0 \\   i \sqrt{2} \chi_5^+ \\   \frac{i (1-\chi_5^+)}{\sqrt{2}} \\  \end{array}  \right),~~~
  Z_{16}=\left(  \begin{array}{c}  i \sqrt{2} \chi_5^- \\  \frac{i (1-\chi_5^-)}{\sqrt{2}} \\  0 \\  0 \\  \end{array}  \right). .
 \label{2docta}
\end{eqnarray}
In terms of~\eqref{2docta} the $20$ cross-ratios then take the values
\begin{flushleft}
 \beqa
u_1&=& \frac{\left(\chi_1{}^-+1\right)
   \left(\chi_2{}^++1\right)}{\left(\chi_1{}^--\chi_2{}^-\right)
   \chi_2{}^+}, \quad u_2 \frac{\left(\chi_1{}^--\chi_2{}^-\right)
   \chi_2{}^+ \left(\chi_3{}^++1\right)}{\left(\chi_1{}^--\chi_3{}^-\right)
   \left(\chi_2{}^++1\right) \chi_3{}^+}, \nonumber \\
 u_3&=& \frac{\left(\chi_1{}^--\chi_3{}^-\right)
   \chi_3{}^+ \left(\chi_4{}^++1\right)}{\left(\chi_1{}^--\chi_4{}^-\right)
   \left(\chi_3{}^++1\right) \chi_4{}^+},\quad u_4= \frac{\chi_2{}^-
   \left(\chi_1{}^--\chi_3{}^-\right) \left(\chi_1{}^+-\chi_2{}^+\right)
   \chi_3{}^+}{\left(\chi_1{}^--\chi_2{}^-\right)
   \chi_3{}^- \chi_2{}^+
   \left(\chi_1{}^+-\chi_3{}^+\right)}, \nonumber \\
 u_5&=& \frac{\chi_3{}^-
   \left(\chi_1{}^--\chi_4{}^-\right) \left(\chi_1{}^+-\chi_3{}^+\right)
   \chi_4{}^+}{\left(\chi_1{}^--\chi_3{}^-\right)
   \chi_4{}^- \chi_3{}^+
   \left(\chi_1{}^+-\chi_4{}^+\right)}, \quad u_6= \frac{\chi_4{}^-
   \left(\chi_1{}^--\chi_5{}^-\right) \left(\chi_1{}^+-\chi_4{}^+\right)
   \chi_5{}^+}{\left(\chi_1{}^--\chi_4{}^-\right)
   \chi_5{}^- \chi_4{}^+
   \left(\chi_1{}^+-\chi_5{}^+\right)}, \nonumber \\
 u_7&=& \frac{\left(\chi_3{}^-+1\right)
   \chi_4{}^- \left(\chi_1{}^+-\chi_4{}^+\right)}{\chi_3{}^-
   \left(\chi_4{}^-+1\right)
   \left(\chi_1{}^+-\chi_3{}^+\right)}, \quad u_8= \frac{\left(\chi_4{}^-+1\right)
   \chi_5{}^- \left(\chi_1{}^+-\chi_5{}^+\right)}{\chi_4{}^-
   \left(\chi_5{}^-+1\right)
   \left(\chi_1{}^+-\chi_4{}^+\right)}, \nonumber \\
 u_9&=& \frac{\left(\chi_5{}^-+1\right)
   \left(\chi_1{}^++1\right)}{\chi_5{}^-
   \left(\chi_1{}^+-\chi_5{}^+\right)}, \quad u_{10}= \frac{\left(\chi_2{}^--\chi_4{}^-\right)
   \left(\chi_5{}^-+1\right)
   \left(\chi_2{}^+-\chi_4{}^+\right)}{\left(\chi_4{}^-+1\right) \left(\chi_2{}^--\chi_5{}^-\right)
   \left(\chi_2{}^+-\chi_5{}^+\right)}, \nonumber \\
 u_{11}&=& -\frac{\left(\chi_2{}^--\chi_5{}^-\right)
   \left(\chi_2{}^+-\chi_5{}^+\right)}{\left(\chi_5{}^-+1\right) \left(\chi_2{}^++1\right)},\quad u_{12}= \frac{\left(\chi_3{}^--\chi_5{}^-\right)
   \left(\chi_2{}^++1\right)
   \left(\chi_3{}^+-\chi_5{}^+\right)}{\left(\chi_2{}^--\chi_5{}^-\right)
   \left(\chi_3{}^++1\right)
   \left(\chi_2{}^+-\chi_5{}^+\right)}, \nonumber \\
 u_{13}&=& \frac{\left(\chi_1{}^--\chi_5{}^-\right)
   \left(\chi_1{}^++1\right) \chi_5{}^+}{\chi_5{}^-
   \left(\chi_1{}^+-\chi_5{}^+\right)}, \quad u_{14}=\frac{\left(\chi_1{}^-+1\right) \left(\chi_1{}^++1\right)}{\chi_1{}^-
   \chi_1{}^+}, \nonumber \\
 u_{15}&=& \frac{\left(\chi_1{}^-+1\right) \chi_2{}^-
   \left(\chi_1{}^+-\chi_2{}^+\right)}{\left(\chi_1{}^--\chi_2{}^-\right)
   \chi_2{}^+}, \quad u_{16}= \frac{\chi_2{}^-
   \left(\chi_3{}^-+1\right)
   \left(\chi_1{}^+-\chi_2{}^+\right)}{\left(\chi_2{}^-+1\right) \chi_3{}^-
   \left(\chi_1{}^+-\chi_3{}^+\right)}, \nonumber \\
 u_{17}&=& \frac{\left(\chi_3{}^-+1\right) \left(\chi_2{}^--\chi_4{}^-\right)
   \left(\chi_2{}^+-\chi_4{}^+\right)}{\left(\chi_2{}^--\chi_3{}^-\right)
   \left(\chi_4{}^-+1\right)
   \left(\chi_2{}^+-\chi_3{}^+\right)}, \quad u_{18}=\frac{\left(\chi_2{}^--\chi_4{}^-\right)
   \left(\chi_3{}^--\chi_5{}^-\right) \left(\chi_2{}^+-\chi_4{}^+\right)
   \left(\chi_3{}^+-\chi_5{}^+\right)}{\left(\chi_3{}^--\chi_4{}^-\right)
   \left(\chi_2{}^--\chi_5{}^-\right) \left(\chi_3{}^+-\chi_4{}^+\right)
   \left(\chi_2{}^+-\chi_5{}^+\right)}, \nonumber \\
 u_{19}&=& -\frac{\left(\chi_3{}^--\chi_5{}^-\right)
   \left(\chi_4{}^++1\right)
   \left(\chi_3{}^+-\chi_5{}^+\right)}{\left(\chi_3{}^--\chi_4{}^-\right)
   \left(\chi_2{}^--\chi_5{}^-\right) \left(\chi_3{}^+-\chi_4{}^+\right)
   \left(\chi_2{}^+-\chi_5{}^+\right)}, \quad u_{20}=\frac{\left(\chi_1{}^--\chi_5{}^-\right)
   \left(\chi_4{}^++1\right) \chi_5{}^+}{\left(\chi_1{}^--\chi_4{}^-\right)
   \chi_4{}^+ \left(\chi_5{}^++1\right)}.
\eeqa

\end{flushleft}

\subsection{A $\Gamma$-function parameterization}
\label{gammas}

Upon expressing the $\delta_{ij}$ in terms of 20 independent variables $c_i$ according to the labeling of the 20 cross-ratios in the previous subsection, the product $\prod_{i<j}^8 \Gamma(\delta_{ij})$ appearing in~\reef{tripleboxone} becomes
\beqa
	\prod_{i<j}^8 \Gamma(\delta_{ij})&=&
	\Gamma \left(c_2-c_3-c_4+c_5+1\right) \Gamma \left(c_5-c_6-c_7+c_8+1\right)\Gamma \left(c_8-c_9-c_{10}+c_{11}+1\right)\nonumber \\ && \Gamma
   \left(-c_1+c_2+c_{11}-c_{12}+1\right) \Gamma \left(c_{13}\right) \Gamma \left(c_6-c_8+c_9+c_{13}\right) \Gamma \left(-c_9-c_{13}-c_{14}\right)\nonumber \\
   && \Gamma
   \left(c_{14}\right) \Gamma \left(c_1+c_9-c_{11}+c_{14}\right) \Gamma \left(-c_1-c_{14}-c_{15}\right) \Gamma \left(c_{15}\right) \Gamma
   \left(c_1-c_2+c_4+c_{15}\right)\nonumber \\
   && \Gamma \left(-c_4-c_{15}-c_{16}\right) \Gamma \left(c_{16}\right) \Gamma \left(c_4-c_5+c_7+c_{16}\right) \Gamma
   \left(-c_7-c_{16}-c_{17}\right) \Gamma \left(c_{17}\right)\nonumber \\
   && \Gamma \left(c_7-c_8+c_{10}+c_{17}\right) \Gamma \left(-c_{10}-c_{17}-c_{18}\right) \Gamma
   \left(c_{18}\right) \Gamma \left(c_{10}-c_{11}+c_{12}+c_{18}\right)\nonumber \\
   && \Gamma \left(-c_{12}-c_{18}-c_{19}\right) \Gamma \left(c_{19}\right) \Gamma
   \left(-c_2+c_3+c_{12}+c_{19}\right) \Gamma \left(-c_6-c_{13}-c_{20}\right)\nonumber \\
   &&\Gamma \left(-c_3-c_{19}-c_{20}\right) \Gamma \left(c_{20}\right) \Gamma
   \left(c_3-c_5+c_6+c_{20}\right).
\eeqa

\end{document}